\documentclass[review]{elsarticle}

\usepackage{lineno,hyperref}

\journal{New Astronomy}

\biboptions{authoryear}

\begin{document}

\begin{frontmatter}

\title{Accurate absolute parameters of the binary system V4089~Sgr\tnoteref{mytitlenote}}


\author[]{M. E. Veramendi}\corref{mycorrespondingauthor}
\ead{mveramendi@icate-conicet.gob.ar, Tel: 54 (0) 264 4213693}
\cortext[mycorrespondingauthor]{Corresponding author}
\author[]{J. F. Gonz\'alez}
\address{Instituto de Ciencias Astron\'omicas, de la Tierra y del Espacio (CONICET-UNSJ), Casilla de Correo 49, 5400 San Juan, Argentina}




\begin{abstract}
We carried out a spectroscopic and photometric analysis of the binary V4089~Sgr with the aim to obtain absolute masses and radii of the components and 
to compare these parameters with stellar evolution theoretical models. We took high-resolution spectra and measured radial velocity using 
standard cross-correlations and a technique of spectral disentangling. Absolute parameters of the components were determined through the 
simultaneous fitting of measured radial velocities and Geneva photometric data available in the literature. In this way we obtained $M_A= 2.584\pm0.008\:{\rm M_{\odot}}$, $M_B= 1.607\pm0.007\:{\rm M_{\odot}}$, $R_A= 3.959\pm0.013\:{\rm R_{\odot}}$, and $R_B= 1.605\pm0.016\:{\rm R_{\odot}}$. The comparison of these parameters 
with two grids of theoretical models led to estimate narrow ranges of possible values for system metallicity and age. According to circularization 
theory it was not expected that the binary had achieved a circular orbit as a result of tidal friction, so the null eccentricity found is an interesting fact. On the other hand, we measured the projected rotational velocity of both components and we conclude that the primary is not synchronized with orbital motion.
\end{abstract}


\end{frontmatter}


\section{Introduction}

Currently, the spectroscopic and photometric study of eclipsing binary stars is a classic astrophysical technique for the determination of stellar masses, radii and luminosities. However, it is only possible to carry out a useful comparison with theoretical models if the errors of such parameters are below $\sim$ 1-3 $\%$ \citep{2010A&ARv..18...67T}. That is, only the parameters determined with high precision provide information about the aspects of the physics of the stars that are not yet well known. \citet{2010A&ARv..18...67T} carried out the most recent compilation of detached binary systems in which masses and radii of the components have been determined with a precision of 3 $\%$ or better and they found a total of 95 binaries.

The aim of this investigation was to obtain highly accurate masses and radii of the components of the spectroscopic and eclipsing 
binary system V4089 Sgr, and to compare those parameters with theoretical models to establish the evolutionary state of these components. 

V4089~Sgr (HD~184\,035) was classified as single-lined spectroscopic binary by \citet{1961MNRAS.123..183B}, who determined a period $P=4.625\:{\rm d}$ and other orbital elements from the fitting of the radial velocity curve of the primary component. \citet{1978mcts.book.....H} published a spectral type A2V for this star, while other catalogues give spectral types A3III \citep[Bright Star Catalogue,][]{1991bsc..book.....H} or A5 IV-III \citep[Hipparcos,][]{1997ESASP1200.....P}. \citet{1983A&AS...52...13W} classified the system as eclipsing binary on the basis of its Geneva $V$ photometry, and subsequently \citet{1984IBVS.2552....1G} proposed a solution for that light curve. \citet{1997A&A...324..137N} fitted again  the radial velocities published by \citet{1961MNRAS.123..183B} and analysed 
the Geneva light-curves in seven colors ($U, B1, B, B2, V1, V, G$). Adopting the mass of the secondary component they estimated the mass and radius of the primary with errors of 17\% and 9\%, respectively. These errors are too high to carry out a useful comparative analysis with theoretical models. In the present paper we present a high-resolution spectroscopic study which allow to calculate stellar parameters for both components with the required accuracy level.



\section{Observations and radial velocities}

Spectroscopic observations were carried out using the 2.15 m telescope and the bench echelle spectrograph EBASIM at the Complejo Astron\'omico El Leoncito (CASLEO), San Juan, Argentina. We obtained 20 spectra of V4089~Sgr during 11 nights spread over three observing runs in June and July 2006. A first set of observations (4 spectra taken in June 2006) cover the wavelength range 3900--5900~\AA, while most of the observations (other 2 spectra taken in June and 14 taken in July 2006, hereafter set 2)  cover the wavelength range 4937--6940~\AA; in both cases the resolving power was $\mathrm{R}$=28\,000. These observations were reduced by using standard data reduction procedures within the NOAO/IRAF package. The spectra were processed by combining echelle orders, normalizing the combined spectrum, eliminating residual cosmic rays, and applying the velocity heliocentric correction.

We applied the disentangling method for double-lined spectra developed by \citet[hereafter GL06]{2006A&A...448..283G} for computing individual spectra of the components and to measure their radial velocities in our set 2 of observations.
Previously we applied standard cross-correlations through the IRAF task {\it fxcor}, in order to obtain the starting velocities required to apply the iterative method GL06. To carry out the correlations, 
we employed spectral regions with metallic lines. The relative velocity between object and template was measured by fitting a Gaussian to the correlation peak. It is worth mentioning that the GL06 method also employs cross-correlations for the radial velocity measurement, but in this case the object spectrum is the observed spectrum after the subtraction of the spectrum computed for the other component. On the other hand, this method requires all the observed spectra to cover the same wavelength range. We employed only the spectra of set 2 in the separation of the components of V4089~Sgr. It was not possible to apply this technique to the set 1 since it contains few spectra with a poor phase coverage. However, we did use these 4 spectra for the orbital analysis, since they did not show blending of the spectral lines of both components, so the radial velocities measured by cross-correlations are realiable.

To be used as templates, we selected synthetic spectra from the POLLUX database \citep{2010A&A...516A..13P}. We employed spectra with $T_{\rm eff}=8500\:\rm{K}$ and $T_{\rm eff}=7500\:\rm{K}$ for primary and secondary components, respectively. Previously, these theoretical spectra were convolved with rotational profiles in order to improve the similarity with the observed spectra. We applied $v\sin i=29\;{\rm km\:s}^{-1}$ and $v\sin i=18\;{\rm km\:s}^{-1}$ to the templates with higher and lower temperature, respectively.

Using the reconstructed individual spectra of the components we determine spectral types A2-3IV and A7V for the primary and secondary components, respectively.  

\section{Analysis}

We determined orbital and stellar parameters of the components by analysing our radial velocities curves along with  the photometric data published by \citet{1997A&A...324..137N} in the Geneva filters $B1, B, B2, V1, V$, and $G$.
To this aim we used the Wilson \& Devinney code \citep[hereafter WD]{1971ApJ...166..605W, 1979ApJ...234.1054W, 1990ApJ...356..613W}.

We fixed the temperature of the primary component according to the value published by North et al. Since the secondary minimum is a total eclipse, they were able to determine the color index of each component and used them to estimate the effective temperatures through the calibration by \citet{1997A&AS..122...51K}. We also fixed the values of the bolometric albedos, limb darkening, and gravity brightening coefficients. 
To model the limb darkening we used a linear law whose bolometric and monochromatic coefficients were selected from the tables by \citet{1993AJ....106.2096V} and \citet{2003A&A...401..657C}, respectively. As usual for stars with radiative envelopes, we adopted values of 1.0 for the exponents of the gravity brightening law ($g$) and for the bolometric albedos of both components. We fitted the orbital parameters $a, e, \omega_0, V_\gamma, i, T_0, P$ and stellar parameters $T_\mathrm{B}, \Omega_\mathrm{A}, \Omega_\mathrm{B}, q$. 

The masses and radii obtained from a first fitting were used to correct for gravitational reddening the measured radial velocities for both components using the relation ${\rm RV}_{\rm cor}={\rm RV}_{\rm obs}-0.635\:(M/R)\;{\rm km\:s}^{-1}$ \citep{2001AJ....121.2657G}. The gravitational reddening correction of the primary star was lower than that of the secondary by $0.22\;{\rm km\:s}^{-1}$, due its more advanced evolutionary state within the main sequence.

\begin{figure}
   \centering
   \includegraphics[width=0.95\linewidth]{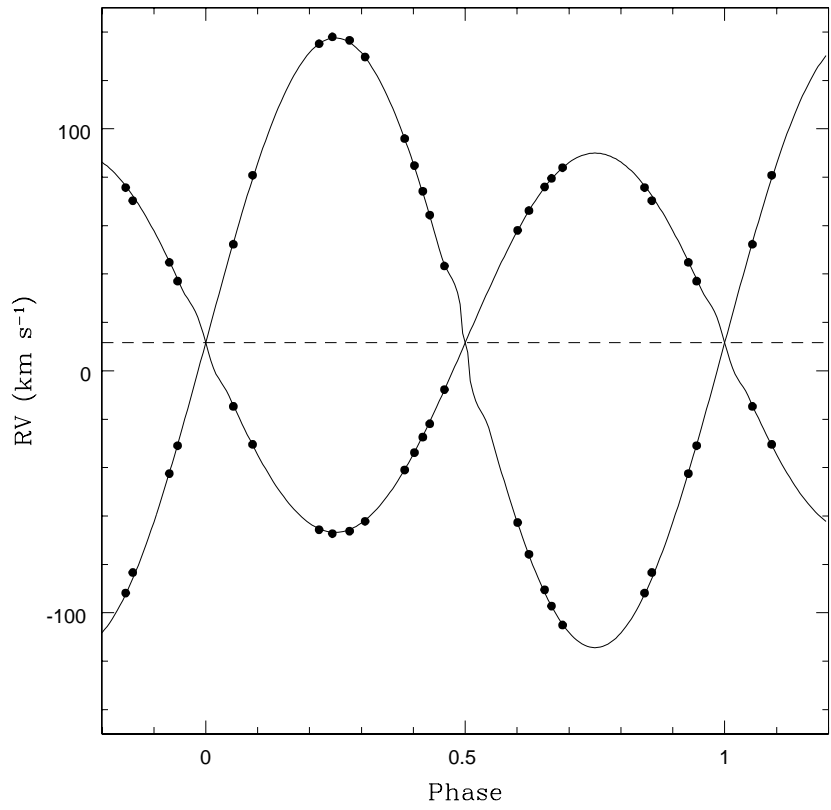}
      \caption{Radial velocity curves of V4089~Sgr. Filled circles represent our measurements. Continuous line shows the orbital fitting. Dashed line indicates the barycentre's velocity.}
         \label{fig1b}
   \end{figure}

\begin{figure}
   \centering
   \includegraphics[width=0.95\linewidth]{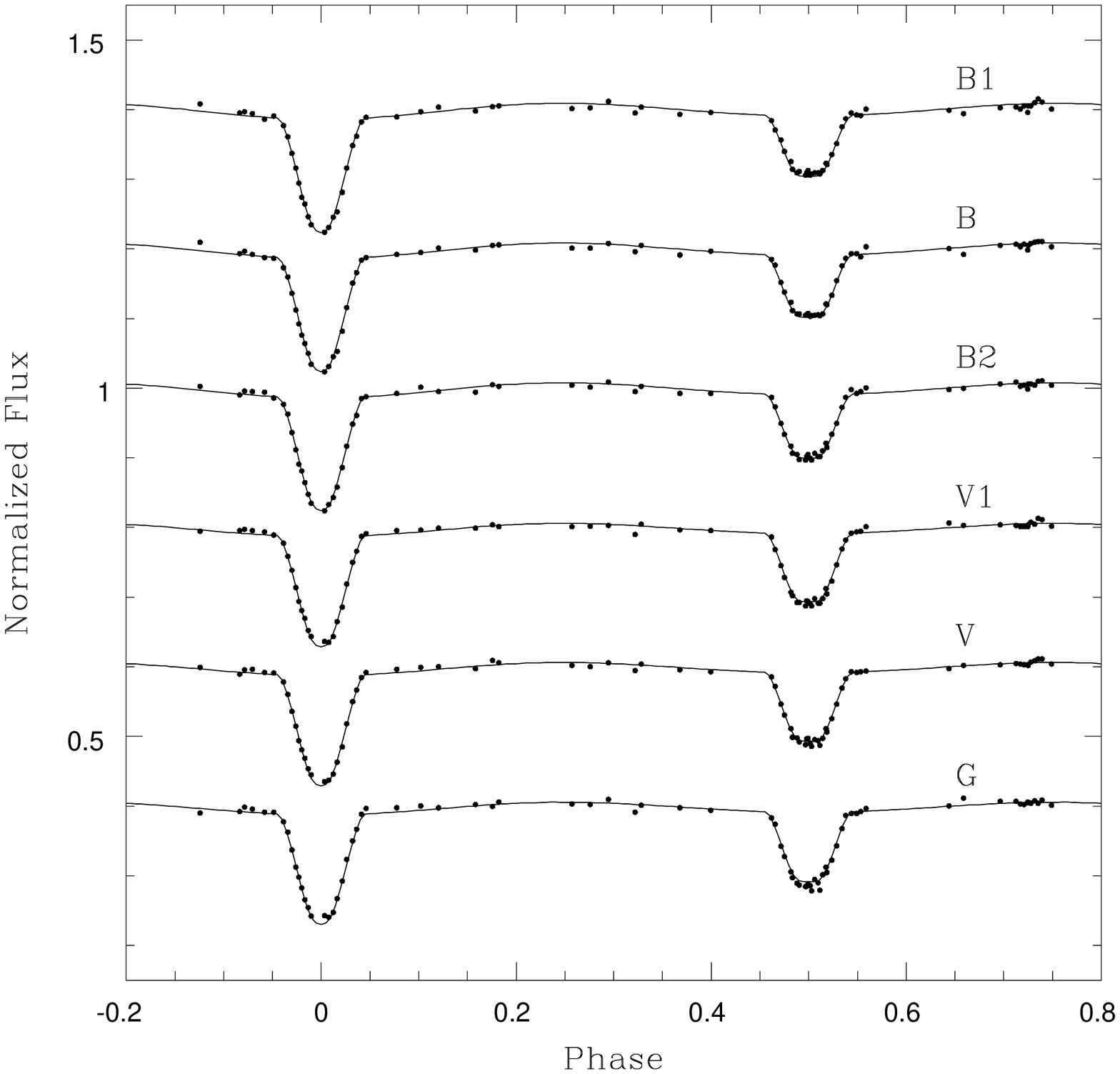}
      \caption{Light curves of V4089~Sgr. Filled circles represent the data by \citet{1997A&A...324..137N} in Geneva filters $B1,B,B2,V1,V,G$, as indicated. Continuous lines show the curve fittings to those data.}
         \label{fig2b}
   \end{figure}

Figure~\ref{fig1b} shows the fitting of the radial velocities of both components. The rms deviations of the radial velocities are $0.6\;{\rm km\:s}^{-1}$ and $0.8\;{\rm km\:s}^{-1}$ for primary and secondary components, respectively. Figure~\ref{fig2b} presents the fitting of the light curves published by North et al. The rms deviations of the data are 0.004 (in normalized flux) for curves $B, B2, V1,\:{\rm and}\:V$ and 0.005 for curves $B1\:{\rm and}\:G$.

Table~\ref{table:1} presents the orbital parameters of the binary V4089~Sgr. We obtained a circular orbit and
 therefore fixed eccentricity at zero in the final calculation of orbital parameters. Errors in $a, i, V_{\gamma}, T_0$, and $P$ are the standard deviations given by WD. To obtain the semiamplitudes $K_A$ and $K_B$ and their corresponding errors we fitted the spectroscopic orbit using the least squares method. 

Table~\ref{table:2} lists the physical parameters obtained for both components. The errors in surface potentials and relative radii are the standard deviations given by WD. We adopted the value given by North et al. for the error in temperature of the primary component and we considered the contribution of that uncertainty to compute the error in secondary temperature. Errors in masses were obtained on the basis of the errors in $K_A, K_B$, and $i$, since we fixed $e=0$ and the uncertainty in period was negligible. We estimated errors in absolute radii by comparing separate solutions obtained for each light curve.

\begin{table*}
\caption{Orbital parameters obtained from the simultaneous fitting of light and radial velocity curves.}
\label{table:1}
\centering
\begin{tabular}{c c} 
\\
\hline
 Parameter & Value \\ 
\hline 
  $P\:({\rm days})$        & $4.6270956\pm0.0000004$ \\    
  $T_{\rm 0,\:Min\:I}\:({\rm HJD-2\,400\,000})$  & $53921.4476\pm0.0006$   \\  
  $V_{\gamma}\:({\rm km\:s}^{-1})$ & $11.60\pm0.06$   \\
  $a\:{\rm(R_{\odot})}$    & $18.821\pm0.016$  \\
  $e$                      & 0 (fixed) \\
  $i\:(^{\circ})$          &  $83.48\pm0.06$  \\
  $K_{\rm A}\:({\rm km\:s}^{-1})$ & $78.4\pm0.3$   \\
  $K_{\rm B}\:({\rm km\:s}^{-1})$ & $125.9\pm0.4$   \\ 
\hline
\end{tabular} 
\end{table*} 

\begin{table*}
\caption{Physical parameters.}
\label{table:2}
\centering
\begin{tabular}{c c c} 
\\
\hline
 Parameter & Primary & Secondary\\ 
\hline 
$T_{\rm eff}\:(\rm K)$ & $8433\pm97$ & $7361\pm105$ \\  
$\Omega$ & $5.401\pm0.008$     & $8.487\pm0.010$  \\ 
$r_{\rm pole}$ & $0.2087\pm0.0008$  & $0.0851\pm0.0003$  \\
$r_{\rm point}$ & $0.2125\pm0.0009$  & $0.0853\pm0.0003$  \\
$r_{\rm side}$ & $0.2103\pm0.0009$  & $0.0852\pm0.0003$  \\
$r_{\rm back}$ & $0.2119\pm0.0009$  & $0.0853\pm0.0003$  \\
$M ({\rm M}_{\odot})$ & $2.584\pm0.008$  & $1.607\pm0.007$   \\
$R_{\rm vol} ({\rm R}_{\odot})$ & $3.959\pm0.013$  & $1.605\pm0.016$  \\
\hline
\end{tabular} 
\end{table*}

We measured projected rotational velocities for both companions applying the method by \citet{2011A&A...531A.143D},
which is suitable for spectra with line blends.
We used 9 spectral windows of the separated spectra obtaining 29.2 km\,s$^{-1}$ for the primary
 and 18.3 km\,s$^{-1}$ for the secondary.
Internal measurements errors, which are due mainly to the spectrum shot noise and
object-template mismatch  are 0.2-0.3 km\,s$^{-1}$. Considering a more realistic estimation, including
uncertainties related with model assumptions (star shape, limb-darkening treatment),
we adopted $(v\sin i)_\mathrm{A}=29.2\pm 0.7$  km\,s$^{-1}$ and $(v\sin i)_\mathrm{B}=18.3\pm 0.5$ km\,s$^{-1}$. 
 
\section{Discussion}

From the simultaneous fitting of our radial velocity data and the photometry by 
\citet{1997A&A...324..137N} we obtained masses and radii of the components of V4089~Sgr with errors below 1\% (see Table~\ref{table:2}). Stellar parameters determined with this accuracy level 
for a binary with very dissimilar components are particularly useful for testing theoretical models.

\begin{figure}
   \centering
   \includegraphics[width=10cm,height=12cm,angle=270]{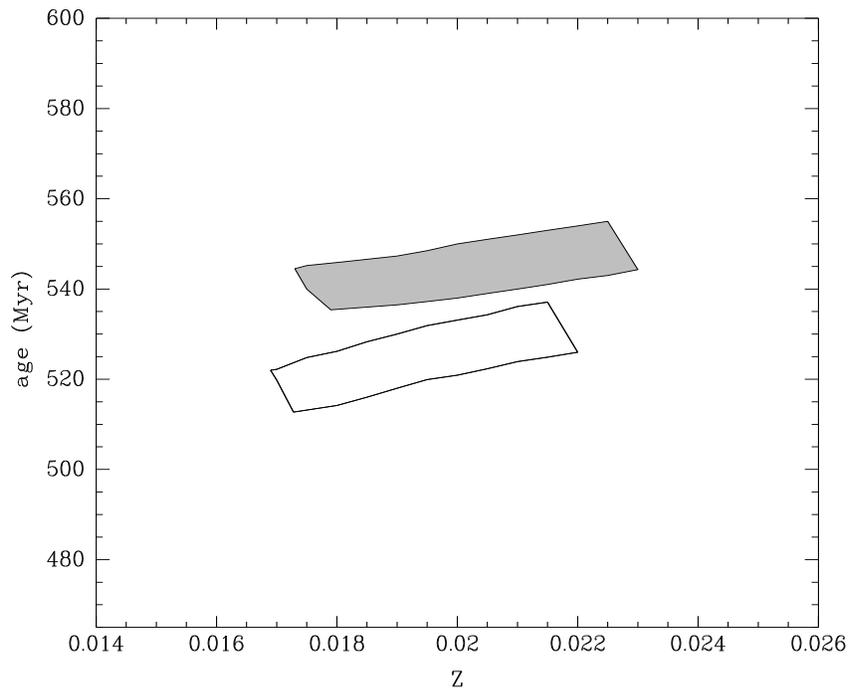}   
      \caption{Possible metallicities and ages of the system V4089~Sgr, according to the Yonsai-Yale \citep{2001ApJS..136..417Y,2004ApJS..155..667D} (grey shaded region) and PARSEC \citep{2012MNRAS.427..127B} (white region) models.}
        \label{figzt}
   \end{figure}

\begin{figure}
   \centering
   \includegraphics[width=10cm,height=12cm,angle=270]{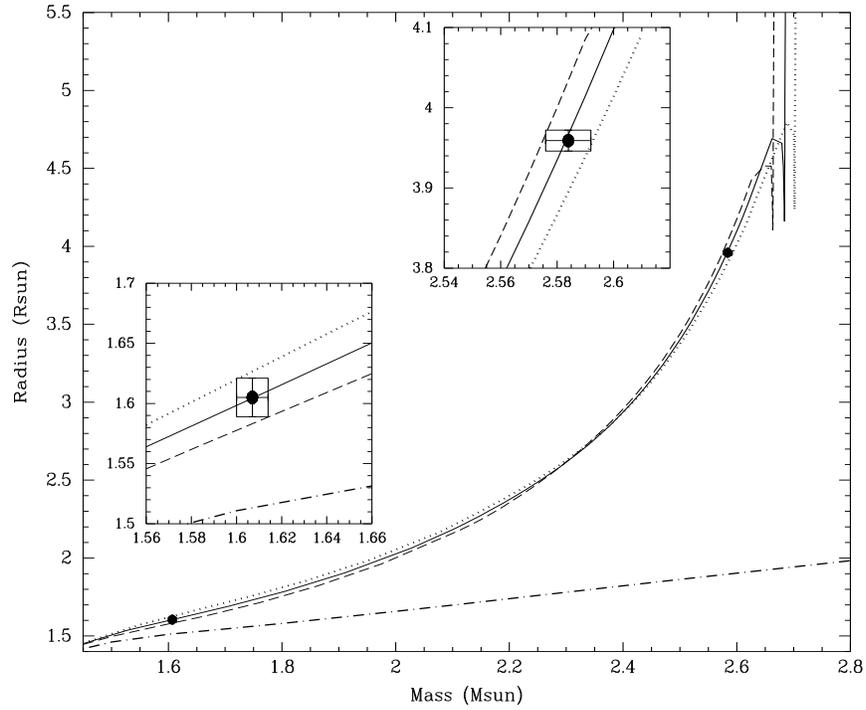}   
      \caption{Comparison of the masses and radii of the components of V4089~Sgr with Yonsai-Yale theoretical isochrones \citep{2001ApJS..136..417Y,2004ApJS..155..667D} for different metallicities. Continuous line shows the best-fitting isochrone for Z=0.0201, corresponding to $\tau=544\:{\rm Myr}$. Dashed and dotted lines represent isochrones for Z=0.0172 and Z=0.0230, respectively, corresponding to the same age. Filled circles with error bars indicate the binary components. Dot-dashed line shows the location of the ZAMS.}
        \label{figyy}
   \end{figure}

\begin{figure}
   \centering
   \includegraphics[width=10cm,height=12cm,angle=270]{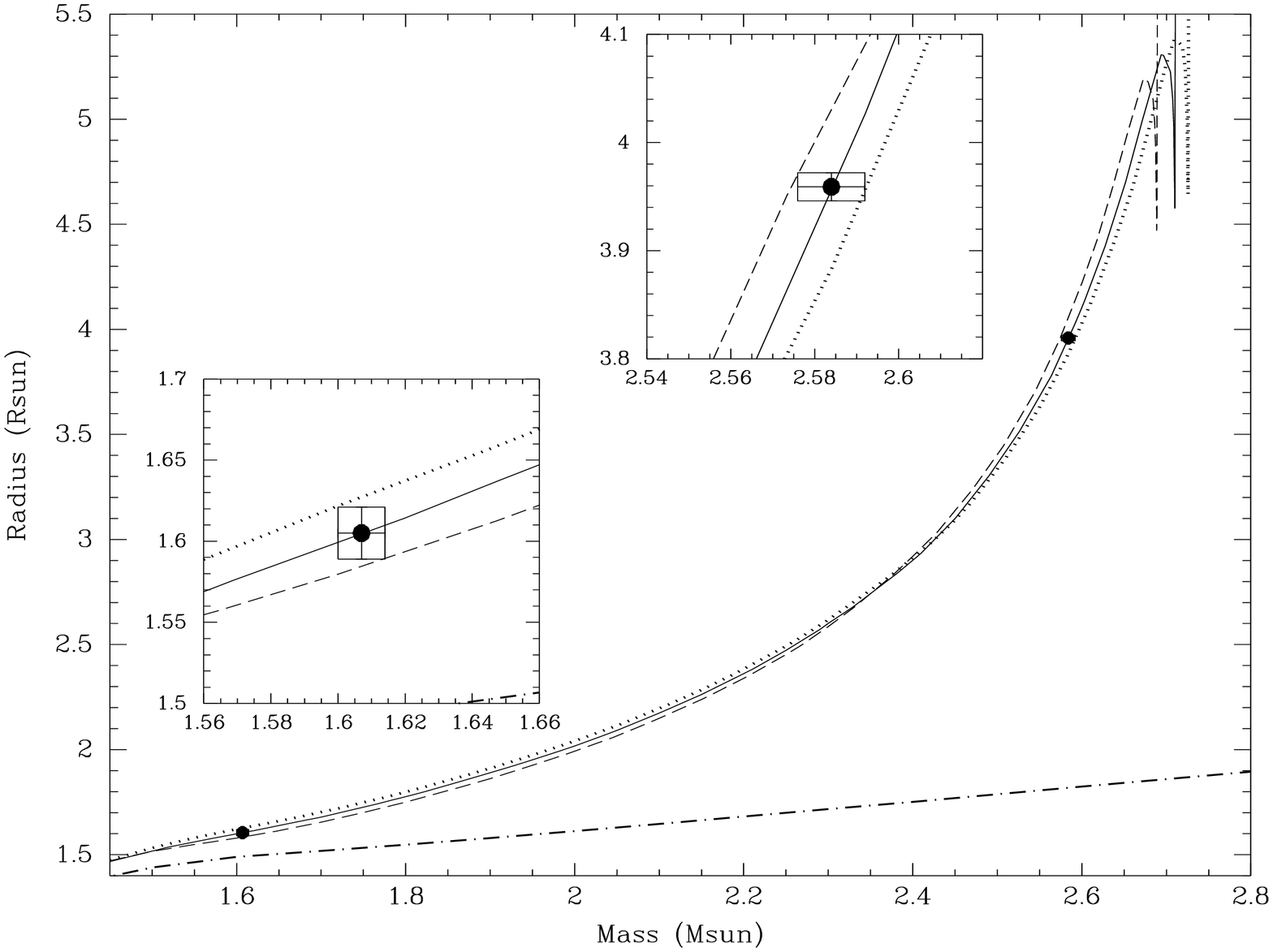}   
      \caption{Comparison of the masses and radii of the components of V4089~Sgr with PARSEC models \citep{2012MNRAS.427..127B} for different metallicities. Continuous line shows the best-fitting isochrone for Z=0.0195, corresponding to $\tau=525\:{\rm Myr}$. Dashed and dotted lines represent isochrones for Z=0.0169 ($\tau=522\:{\rm Myr}$) and Z=0.0220 ($\tau=526\:{\rm Myr}$), respectively. Filled circles with error bars indicate the binary components. Dot-dashed line shows the location of the ZAMS.}
        \label{figparsec}
   \end{figure}

We compared the obtained masses and radii with the Yonsai-Yale \citep{2001ApJS..136..417Y,2004ApJS..155..667D} and PARSEC \citep{2012MNRAS.427..127B} theoretical evolutionary models, taking advantage of the algorithms given by the authors to interpolate in age and metallicity. In both cases, we first refined the original isochrone grids and then we employed the denser new grids to study the possible values of metallicity and age of the binary according to each theoretical model. To this end we selected all isochrones that agree, within observational errors, with the masses and radii of both components. The results are showed in Figure~\ref{figzt}.
PARSEC models give a younger age than the Yonsai-Yale models. Specifically, the former indicate an age between 513 and 537 Myr, while the latter lead to the range 535--555 Myr. This age difference could be possibly associated to the different prescriptions of overshooting of these models. On the other hand, both models indicate essentially the same metallicity range, since PARSEC lead to Z between 0.0169 and 0.0220, while Yonsai-Yale lead to Z between 0.0172 and 0.0230. 
The comparison with theoretical models, therefore, has allowed us to determine the primordial metallicity of the system with a
formal uncertainty of about $\pm$0.07 dex, an accuracy difficult to achieve from a standard abundance analysis in high-resolution
spectra. 
Figures~\ref{figyy} and ~\ref{figparsec} illustrate the determination of these narrow metallicity ranges through comparison of the masses and radii obtained with isochrones of different metallicities for Yonsai-Yale and PARSEC models, respectively. 
The comparison of the primary spectrum with solar-composition synthetic templates and with other stars of similar 
spectral types, suggest that its atmospheric abundances are not normal.
In fact, lines of several ions appear with abnormal intensity, Ba\,II and Y\,II being the most notorious, whose lines have
equivalent widths about twice than normal stars of the same temperature.
A detailed chemical analysis of this system would be valuable for the knowledge of chemically peculiar stars, since
the mass and evolutionary state are accurately known, and even the original composition is stringently constrained by the 
stellar parameters. 
For this kind of study, however, a large number of observations in the blue spectral region would be desirable, since
our four blue spectra are not enough for a realiable reconstruction of the primary spectrum in that region.
 
In order to evaluate if the observed null eccentricity is an expected consequence of orbit circularization owing to tidal interaction, we model the evolution of V4089~Sgr using the Binary-Star Evolution (BSE) code developed by \citet{2002MNRAS.329..897H}.  On the basis of the theories and formulations by \citet{1977A&A....57..383Z} and \citet{1981A&A....99..126H}, BSE computes the variations in orbital eccentricity 
and spin of the components. We applied the code to model the evolution of fictitious binaries with different primordial eccentricities, 
assuming the masses of the components of V4089~Sgr and an initial separation that would lead to the observed separation for the present age of 
the system. In no case the orbit circularization by tidal interaction was achieved. Thus, possibly the binary was formed in a circular orbit. 
It is worth mentioning that there are several well studied binaries with circular orbits for which the theory predicts circularization 
time-scales greater than their present ages \citep{2010MNRAS.401..257K}.

An additional argument supporting the idea that the orbit has not been circularized by tidal interactions
is the fact that the stellar rotation is not synchronized with the orbital motion. In standard models of
tidal evolution, the synchronization of stellar rotation is achieved well before orbit circularization \citep{1981A&A....99..126H}.
Assuming that rotational axes are perpendicular to the orbital plane, 
rotation periods can be determined from  projected rotational velocities and  stellar radii: 
$P_\mathrm{rot}= 2\pi R \sin i/ (v\sin i)$.
For the primary and secondary of V4089~Sgr we found $P_\mathrm{rot,A} = 6.86\pm 0.16$ d and
$P_\mathrm{rot,B} = 4.41\pm 0.13$ d. The primary star is rotating clearly subsynchronously.

If stellar rotation is not locked to orbital motion by tidal forces, the low rotational velocity observed 
offers a test for the models of rotational evolution of intermediate-mass stars.
The variation of the surface angular velocity during the main-sequence phase
depends not only on the changes of the moment of inertia of the star, but also on the efficiency
of the angular momentum transport in the stellar interior and the angular momentum loss
by stellar winds \citep{2013A&A...553A..24G}. 
Eventually the contribution of magnetic braking might be significant even in early-type stars
\citep{2011A&A...525L..11M}. 
In particular, a decrease of the surface angular velocity is expected if the internal transport
is not efficient and the outer layers conserve its specific angular momentum as the star expands.
Theoretical models predict in general a moderate decreasing of the surface rotational velocity during the 
main-sequence stage \citep{2013A&A...553A..24G}. 
From the observational point of view, \citet{2012A&A...537A.120Z} found evidence 
that the surface rotational velocity of intermediate-mass stars increases during
the first third of the main-sequence and decreases the last third.
Being the primary component of V4089\,Sgr in an advanced evolutionary state within the
main-sequence, its low rotational velocity would be a reasonable result 
if its evolution has not been affected significantly by binary tides.


Given the difference between the components, the primary component is close to the TAMS (terminal-age main sequence), having passed approximately the $92\%$ of its life on the Main-Sequence, while the secondary is still close to the ZAMS (zero-age main sequence)        
(see Figures~\ref{figyy} and ~\ref{figparsec}). 
The primary star is expected to fills its Roche lobe when it reaches a radius of $7.91\,{\rm R_{\odot}}$,  which will occur, according to Yonsai-Yale models, at an age of 594 Myr, when this component is crossing the Hertzsprung gap of the HR diagram. At that time, the secondary  will have passed only $21\%$ of its life in the main sequence. 
Once the primary fills its Roche lobe, it will begin to transfer mass to its companion in a type-B mass transfer process. 
According to BSE, when the primary reaches the First Giant Branch (about 1 Myr after the beginning of the Roche lobe overflow), the binary will reach a common-envelope state and both components will coalesce forming a new single giant star. 

\section*{Acknowledgements}

This research was achieved using the POLLUX database (http://pollux.graal.univ-montp2.fr/), 
operated at LUPM (Universit\'e Montpellier II - CNRS, France) with the support of the PNPS and INSU.




\section*{References}

\bibliography{biblio}

\begin{thebibliography}{28}
\expandafter\ifx\csname natexlab\endcsname\relax\def\natexlab#1{#1}\fi
\providecommand{\url}[1]{\texttt{#1}}
\providecommand{\href}[2]{#2}
\providecommand{\path}[1]{#1}
\providecommand{\DOIprefix}{doi:}
\providecommand{\ArXivprefix}{arXiv:}
\providecommand{\URLprefix}{URL: }
\providecommand{\Pubmedprefix}{pmid:}
\providecommand{\doi}[1]{\href{http://dx.doi.org/#1}{\path{#1}}}
\providecommand{\Pubmed}[1]{\href{pmid:#1}{\path{#1}}}
\providecommand{\bibinfo}[2]{#2}
\ifx\xfnm\relax \def\xfnm[#1]{\unskip,\space#1}\fi
\bibitem[{{Bressan} et~al.(2012){Bressan}, {Marigo}, {Girardi}, {Salasnich},
  {Dal Cero}, {Rubele} and {Nanni}}]{2012MNRAS.427..127B}
\bibinfo{author}{{Bressan}, A.}, \bibinfo{author}{{Marigo}, P.},
  \bibinfo{author}{{Girardi}, L.}, \bibinfo{author}{{Salasnich}, B.},
  \bibinfo{author}{{Dal Cero}, C.}, \bibinfo{author}{{Rubele}, S.},
  \bibinfo{author}{{Nanni}, A.}, \bibinfo{year}{2012}.
\newblock \bibinfo{title}{{PARSEC: stellar tracks and isochrones with the
  PAdova and TRieste Stellar Evolution Code}}.
\newblock \bibinfo{journal}{MNRAS} \bibinfo{volume}{427},
  \bibinfo{pages}{127--145}.
\newblock \DOIprefix\doi{10.1111/j.1365-2966.2012.21948.x},
  \href{http://arxiv.org/abs/1208.4498}{\tt arXiv:1208.4498}.
\bibitem[{{Buscombe} and {Morris}(1961)}]{1961MNRAS.123..183B}
\bibinfo{author}{{Buscombe}, W.}, \bibinfo{author}{{Morris}, P.M.},
  \bibinfo{year}{1961}.
\newblock \bibinfo{title}{{Three southern spectroscopic binaries}}.
\newblock \bibinfo{journal}{MNRAS} \bibinfo{volume}{123}, \bibinfo{pages}{183}.
\bibitem[{{Claret}(2003)}]{2003A&A...401..657C}
\bibinfo{author}{{Claret}, A.}, \bibinfo{year}{2003}.
\newblock \bibinfo{title}{{A new non-linear limb-darkening law for LTE stellar
  atmosphere models II K at several surface gravities. Geneva and Walraven
  systems: Calculations for -5.0 $\le$ log [M/H] $\le$ +1 2000 K $\le$
  T$_{eff}$ $\le$ 50,000 K at several surface gravities}}.
\newblock \bibinfo{journal}{A\&A} \bibinfo{volume}{401},
  \bibinfo{pages}{657--660}.
\newblock \DOIprefix\doi{10.1051/0004-6361:20030142}.
\bibitem[{{Demarque} et~al.(2004){Demarque}, {Woo}, {Kim} and
  {Yi}}]{2004ApJS..155..667D}
\bibinfo{author}{{Demarque}, P.}, \bibinfo{author}{{Woo}, J.H.},
  \bibinfo{author}{{Kim}, Y.C.}, \bibinfo{author}{{Yi}, S.K.},
  \bibinfo{year}{2004}.
\newblock \bibinfo{title}{{Y$^{2}$ Isochrones with an Improved Core Overshoot
  Treatment}}.
\newblock \bibinfo{journal}{ApJS} \bibinfo{volume}{155},
  \bibinfo{pages}{667--674}.
\newblock \DOIprefix\doi{10.1086/424966}.
\bibitem[{{D{\'{\i}}az} et~al.(2011){D{\'{\i}}az}, {Gonz{\'a}lez}, {Levato} and
  {Grosso}}]{2011A&A...531A.143D}
\bibinfo{author}{{D{\'{\i}}az}, C.G.}, \bibinfo{author}{{Gonz{\'a}lez}, J.F.},
  \bibinfo{author}{{Levato}, H.}, \bibinfo{author}{{Grosso}, M.},
  \bibinfo{year}{2011}.
\newblock \bibinfo{title}{{Accurate stellar rotational velocities using the
  Fourier transform of the cross correlation maximum}}.
\newblock \bibinfo{journal}{A\&A} \bibinfo{volume}{531}, \bibinfo{pages}{A143}.
\newblock \DOIprefix\doi{10.1051/0004-6361/201016386},
  \href{http://arxiv.org/abs/1012.4858}{\tt arXiv:1012.4858}.
\bibitem[{{Gaspani}(1984)}]{1984IBVS.2552....1G}
\bibinfo{author}{{Gaspani}, A.}, \bibinfo{year}{1984}.
\newblock \bibinfo{title}{{Preliminary Orbital Elements of HD 184035}}.
\newblock \bibinfo{journal}{Information Bulletin on Variable Stars}
  \bibinfo{volume}{2552}, \bibinfo{pages}{1}.
\bibitem[{{Georgy} et~al.(2013){Georgy}, {Ekstr{\"o}m}, {Granada}, {Meynet},
  {Mowlavi}, {Eggenberger} and {Maeder}}]{2013A&A...553A..24G}
\bibinfo{author}{{Georgy}, C.}, \bibinfo{author}{{Ekstr{\"o}m}, S.},
  \bibinfo{author}{{Granada}, A.}, \bibinfo{author}{{Meynet}, G.},
  \bibinfo{author}{{Mowlavi}, N.}, \bibinfo{author}{{Eggenberger}, P.},
  \bibinfo{author}{{Maeder}, A.}, \bibinfo{year}{2013}.
\newblock \bibinfo{title}{{Populations of rotating stars. I. Models from 1.7 to
  15 M$_{⊙}$ at Z = 0.014, 0.006, and 0.002 with
  {$\Omega$}/{$\Omega$}$_{crit}$ between 0 and 1}}.
\newblock \bibinfo{journal}{A\&A} \bibinfo{volume}{553}, \bibinfo{pages}{A24}.
\newblock \DOIprefix\doi{10.1051/0004-6361/201220558},
  \href{http://arxiv.org/abs/1303.2321}{\tt arXiv:1303.2321}.
\bibitem[{{Gonz{\'a}lez} and {Lapasset}(2001)}]{2001AJ....121.2657G}
\bibinfo{author}{{Gonz{\'a}lez}, J.F.}, \bibinfo{author}{{Lapasset}, E.},
  \bibinfo{year}{2001}.
\newblock \bibinfo{title}{{Radial Velocities and Kinematic Membership in the
  Open Cluster NGC 3114}}.
\newblock \bibinfo{journal}{AJ} \bibinfo{volume}{121},
  \bibinfo{pages}{2657--2663}.
\newblock \DOIprefix\doi{10.1086/320396}.
\bibitem[{{Gonz{\'a}lez} and {Levato}(2006)}]{2006A&A...448..283G}
\bibinfo{author}{{Gonz{\'a}lez}, J.F.}, \bibinfo{author}{{Levato}, H.},
  \bibinfo{year}{2006}.
\newblock \bibinfo{title}{{Separation of composite spectra: the spectroscopic
  detection of an eclipsing binary star}}.
\newblock \bibinfo{journal}{A\&A} \bibinfo{volume}{448},
  \bibinfo{pages}{283--292}.
\newblock \DOIprefix\doi{10.1051/0004-6361:20053177}.
\bibitem[{{Hoffleit} and {Jaschek}(1991)}]{1991bsc..book.....H}
\bibinfo{author}{{Hoffleit}, D.}, \bibinfo{author}{{Jaschek}, C..},
  \bibinfo{year}{1991}.
\newblock \bibinfo{title}{{The Bright star catalogue}}.
\bibitem[{{Houk}(1978)}]{1978mcts.book.....H}
\bibinfo{author}{{Houk}, N.}, \bibinfo{year}{1978}.
\newblock \bibinfo{title}{{Michigan catalogue of two-dimensional spectral types
  for the HD stars}}.
\bibitem[{{Hurley} et~al.(2002){Hurley}, {Tout} and
  {Pols}}]{2002MNRAS.329..897H}
\bibinfo{author}{{Hurley}, J.R.}, \bibinfo{author}{{Tout}, C.A.},
  \bibinfo{author}{{Pols}, O.R.}, \bibinfo{year}{2002}.
\newblock \bibinfo{title}{{Evolution of binary stars and the effect of tides on
  binary populations}}.
\newblock \bibinfo{journal}{MNRAS} \bibinfo{volume}{329},
  \bibinfo{pages}{897--928}.
\newblock \DOIprefix\doi{10.1046/j.1365-8711.2002.05038.x},
  \href{http://arxiv.org/abs/astro-ph/0201220}{\tt arXiv:astro-ph/0201220}.
\bibitem[{{Hut}(1981)}]{1981A&A....99..126H}
\bibinfo{author}{{Hut}, P.}, \bibinfo{year}{1981}.
\newblock \bibinfo{title}{{Tidal evolution in close binary systems}}.
\newblock \bibinfo{journal}{A\&A} \bibinfo{volume}{99},
  \bibinfo{pages}{126--140}.
\bibitem[{{Khaliullin} and {Khaliullina}(2010)}]{2010MNRAS.401..257K}
\bibinfo{author}{{Khaliullin}, K.F.}, \bibinfo{author}{{Khaliullina}, A.I.},
  \bibinfo{year}{2010}.
\newblock \bibinfo{title}{{Synchronization and circularization in early-type
  binaries on main sequence}}.
\newblock \bibinfo{journal}{MNRAS} \bibinfo{volume}{401},
  \bibinfo{pages}{257--274}.
\newblock \DOIprefix\doi{10.1111/j.1365-2966.2009.15630.x}.
\bibitem[{{Kunzli} et~al.(1997){Kunzli}, {North}, {Kurucz} and
  {Nicolet}}]{1997A&AS..122...51K}
\bibinfo{author}{{Kunzli}, M.}, \bibinfo{author}{{North}, P.},
  \bibinfo{author}{{Kurucz}, R.L.}, \bibinfo{author}{{Nicolet}, B.},
  \bibinfo{year}{1997}.
\newblock \bibinfo{title}{{A calibration of Geneva photometry for B to G stars
  in terms of T$_{eff}$, log G and [M/H]}}.
\newblock \bibinfo{journal}{A\&AS} \bibinfo{volume}{122},
  \bibinfo{pages}{51--77}.
\newblock \DOIprefix\doi{10.1051/aas:1997291}.
\bibitem[{{Meynet} et~al.(2011){Meynet}, {Eggenberger} and
  {Maeder}}]{2011A&A...525L..11M}
\bibinfo{author}{{Meynet}, G.}, \bibinfo{author}{{Eggenberger}, P.},
  \bibinfo{author}{{Maeder}, A.}, \bibinfo{year}{2011}.
\newblock \bibinfo{title}{{Massive star models with magnetic braking}}.
\newblock \bibinfo{journal}{A\&A} \bibinfo{volume}{525}, \bibinfo{pages}{L11}.
\newblock \DOIprefix\doi{10.1051/0004-6361/201016017},
  \href{http://arxiv.org/abs/1011.5795}{\tt arXiv:1011.5795}.
\bibitem[{{North} et~al.(1997){North}, {Studer} and
  {Kunzli}}]{1997A&A...324..137N}
\bibinfo{author}{{North}, P.}, \bibinfo{author}{{Studer}, M.},
  \bibinfo{author}{{Kunzli}, M.}, \bibinfo{year}{1997}.
\newblock \bibinfo{title}{{Eclipsing binaries with candidate CP stars. I.
  Parameters of the systems HD 143654, HD 184035 and HD 185257.}}
\newblock \bibinfo{journal}{A\&A} \bibinfo{volume}{324},
  \bibinfo{pages}{137--154}.
\bibitem[{{Palacios} et~al.(2010){Palacios}, {Gebran}, {Josselin}, {Martins},
  {Plez}, {Belmas} and {L{\`e}bre}}]{2010A&A...516A..13P}
\bibinfo{author}{{Palacios}, A.}, \bibinfo{author}{{Gebran}, M.},
  \bibinfo{author}{{Josselin}, E.}, \bibinfo{author}{{Martins}, F.},
  \bibinfo{author}{{Plez}, B.}, \bibinfo{author}{{Belmas}, M.},
  \bibinfo{author}{{L{\`e}bre}, A.}, \bibinfo{year}{2010}.
\newblock \bibinfo{title}{{POLLUX: a database of synthetic stellar spectra}}.
\newblock \bibinfo{journal}{A\&A} \bibinfo{volume}{516}, \bibinfo{pages}{A13}.
\newblock \DOIprefix\doi{10.1051/0004-6361/200913932},
  \href{http://arxiv.org/abs/1003.4682}{\tt arXiv:1003.4682}.
\bibitem[{{Perryman} and {ESA}(1997)}]{1997ESASP1200.....P}
\bibinfo{editor}{{Perryman}, M.A.C.}, \bibinfo{editor}{{ESA}} (Eds.),
  \bibinfo{year}{1997}.
\newblock \bibinfo{title}{{The HIPPARCOS and TYCHO catalogues. Astrometric and
  photometric star catalogues derived from the ESA HIPPARCOS Space Astrometry
  Mission}}. volume \bibinfo{volume}{1200} of \textit{\bibinfo{series}{ESA
  Special Publication}}.
\bibitem[{{Torres} et~al.(2010){Torres}, {Andersen} and
  {Gim{\'e}nez}}]{2010A&ARv..18...67T}
\bibinfo{author}{{Torres}, G.}, \bibinfo{author}{{Andersen}, J.},
  \bibinfo{author}{{Gim{\'e}nez}, A.}, \bibinfo{year}{2010}.
\newblock \bibinfo{title}{{Accurate masses and radii of normal stars: modern
  results and applications}}.
\newblock \bibinfo{journal}{A\&AR} \bibinfo{volume}{18},
  \bibinfo{pages}{67--126}.
\newblock \DOIprefix\doi{10.1007/s00159-009-0025-1},
  \href{http://arxiv.org/abs/0908.2624}{\tt arXiv:0908.2624}.
\bibitem[{{van Hamme}(1993)}]{1993AJ....106.2096V}
\bibinfo{author}{{van Hamme}, W.}, \bibinfo{year}{1993}.
\newblock \bibinfo{title}{{New limb-darkening coefficients for modeling binary
  star light curves}}.
\newblock \bibinfo{journal}{AJ} \bibinfo{volume}{106},
  \bibinfo{pages}{2096--2117}.
\newblock \DOIprefix\doi{10.1086/116788}.
\bibitem[{{Waelkens} and {Rufener}(1983)}]{1983A&AS...52...13W}
\bibinfo{author}{{Waelkens}, C.}, \bibinfo{author}{{Rufener}, F.},
  \bibinfo{year}{1983}.
\newblock \bibinfo{title}{{Light curves of four southern bright hitherto
  unknown eclipsing binaries}}.
\newblock \bibinfo{journal}{A\&AS} \bibinfo{volume}{52},
  \bibinfo{pages}{13--20}.
\bibitem[{{Wilson}(1979)}]{1979ApJ...234.1054W}
\bibinfo{author}{{Wilson}, R.E.}, \bibinfo{year}{1979}.
\newblock \bibinfo{title}{{Eccentric orbit generalization and simultaneous
  solution of binary star light and velocity curves}}.
\newblock \bibinfo{journal}{ApJ} \bibinfo{volume}{234},
  \bibinfo{pages}{1054--1066}.
\newblock \DOIprefix\doi{10.1086/157588}.
\bibitem[{{Wilson}(1990)}]{1990ApJ...356..613W}
\bibinfo{author}{{Wilson}, R.E.}, \bibinfo{year}{1990}.
\newblock \bibinfo{title}{{Accuracy and efficiency in the binary star
  reflection effect}}.
\newblock \bibinfo{journal}{ApJ} \bibinfo{volume}{356},
  \bibinfo{pages}{613--622}.
\newblock \DOIprefix\doi{10.1086/168867}.
\bibitem[{{Wilson} and {Devinney}(1971)}]{1971ApJ...166..605W}
\bibinfo{author}{{Wilson}, R.E.}, \bibinfo{author}{{Devinney}, E.J.},
  \bibinfo{year}{1971}.
\newblock \bibinfo{title}{{Realization of Accurate Close-Binary Light Curves:
  Application to MR Cygni}}.
\newblock \bibinfo{journal}{ApJ} \bibinfo{volume}{166}, \bibinfo{pages}{605}.
\newblock \DOIprefix\doi{10.1086/150986}.
\bibitem[{{Yi} et~al.(2001){Yi}, {Demarque}, {Kim}, {Lee}, {Ree}, {Lejeune} and
  {Barnes}}]{2001ApJS..136..417Y}
\bibinfo{author}{{Yi}, S.}, \bibinfo{author}{{Demarque}, P.},
  \bibinfo{author}{{Kim}, Y.C.}, \bibinfo{author}{{Lee}, Y.W.},
  \bibinfo{author}{{Ree}, C.H.}, \bibinfo{author}{{Lejeune}, T.},
  \bibinfo{author}{{Barnes}, S.}, \bibinfo{year}{2001}.
\newblock \bibinfo{title}{{Toward Better Age Estimates for Stellar Populations:
  The Y$^{2}$ Isochrones for Solar Mixture}}.
\newblock \bibinfo{journal}{ApJS} \bibinfo{volume}{136},
  \bibinfo{pages}{417--437}.
\newblock \DOIprefix\doi{10.1086/321795},
  \href{http://arxiv.org/abs/astro-ph/0104292}{\tt arXiv:astro-ph/0104292}.
\bibitem[{{Zahn}(1977)}]{1977A&A....57..383Z}
\bibinfo{author}{{Zahn}, J.P.}, \bibinfo{year}{1977}.
\newblock \bibinfo{title}{{Tidal friction in close binary stars}}.
\newblock \bibinfo{journal}{A\&A} \bibinfo{volume}{57},
  \bibinfo{pages}{383--394}.
\bibitem[{{Zorec} and {Royer}(2012)}]{2012A&A...537A.120Z}
\bibinfo{author}{{Zorec}, J.}, \bibinfo{author}{{Royer}, F.},
  \bibinfo{year}{2012}.
\newblock \bibinfo{title}{{Rotational velocities of A-type stars. IV. Evolution
  of rotational velocities}}.
\newblock \bibinfo{journal}{A\&A} \bibinfo{volume}{537}, \bibinfo{pages}{A120}.
\newblock \DOIprefix\doi{10.1051/0004-6361/201117691},
  \href{http://arxiv.org/abs/1201.2052}{\tt arXiv:1201.2052}.

\end{thebibliography}

\end{document}